\documentclass[12pt,longbibliography]{iopart}
\begin{document}
\title{{Transition probabilities for a Rydberg atom in the field of a gravitational wave}}
\author{Uwe R. Fischer\,\footnote{Present address: Seoul National University, Department of Physics and Astronomy, Center for Theoretical Physics, Seoul 08826, Korea}}%
\address{Lehr- und Forschungsbereich Theoretische Astrophysik, 
Auf der Morgenstelle 10, 72076 T\"ubingen, Germany}

\date{\today}
             
\begin{abstract}
The possibility  of an atomic detection of gravitational waves on earth
is considered. The combination of extremely high lifetimes
and resulting small radiative transition
probabilities with rapidly growing interaction strength for Rydberg
atoms having principal quantum numbers in a region $10^4\ldots 10^5$ might 
result in transition probabilities which are high enough 
to open up such a possibility.
Transition probabilities and absorption cross sections are calculated as
a function of the relevant quantum numbers of a highly excited electron. 
The orders of magnitude for the transition rate are evaluated for a 
realistic source of gravitational radiation. 
It is shown that no specific particle property enters 
the expression for the 
absorption cross section for gravitational waves.
The only fundamental constant contained in this cross section is, 
apart from the fine structure constant $\alpha$, the 
Planck length $L^*=(\hbar G /c^3)^{1/2}$.
\end{abstract}


\section{Introduction}        
In spite of the domination of electromagnetic forces in an atom effects of the 
curvature of spacetime may be significant either in the case of very strong 
gravitational fields perturbing its energy levels or the resonant quadrupole 
absorption of a gravitational wave.  These interaction processes become
important for Rydberg atoms due to their high principal quantum numbers $n$. 
The reason comes from  the dependence 
of the interaction operator proportional to $r^2$, where $r\propto n^2$ is the 
(averaged) distance between core and excited electron, so that matrix elements
of this operator are proportional to $n^4$.
So, if the typical magnitudes of the effects are great enough
and there exist atoms with the required values of $n$,
the influence either on energy levels or the resonant absorption 
of the waves might principally serve as a gravitational wave detector.
The first of these possibilities was checked by
Leen et. al. (\cite{leen}).
They  considered  the 
remote detection of gravitational waves by calculating the modification of the
spectral lines in strong gravitational wave fields. (An analytical calculation 
for the modification of the levels in strong {\em static} fields was carried 
out by Gill et. al. in \cite{gill} and recently Pinto \cite{pinto}  also
dealt with this subject.)

The examination of the possibility using resonant absorption 
undertaken here originated in the consideration 
of an highly excited electron in an earth laboratory. Principal quantum numbers
which are accessible today by excitation of the atom with
laser-spectroscopic techniques lie in a range $n\sim 500\ldots 1000$. 
Correspondingly a perfect isolation of the ion trap, 
in which the  atom is caught, against electromagnetic radiation of lower
frequency  is within reach.
If one calculates the 
quantum numbers for a transition between adjacent principal quantum numbers within
a frequency region $10\ldots 1000$ Hz typical for a {\em pulsar} source 
one obtains
$n\sim 10^4\ldots 10^5$.
The energy levels for such highly excited atoms are defined so extremely sharp  
(corresponding to (theoretical) lifetimes up to ten million years ($n\sim 10^5$))
that nearly every field  quantum would be able to cause a 
transition, so that {\em a priori} it is not obvious
to deny an earth-based detection in this way if one day atoms with the 
required values of $n$ may be excited and held stable. The combination of 
extremely high lifetimes ($\propto n^5$, see section \ref{depend})
and rapid growing of the interaction strength ($\propto n^4$)
might result in an overall effect which cannot be neglected.

In the calculation performed in this paper
we are concerned with the problem of an interaction  of highly 
excited atomic systems with monochromatic gravitational waves. 
That the examination of such interaction-processes makes any physical sense 
only for higher excited atoms at least within the   classical methods of quantum theory  
becomes clear by an inspection of the magnitude of
vacuum fluctuations of the gravitational field.
The spacetime-structure looses its classical-deterministic character on 
scales $L$ to an extent given by $\Delta g=L/L^*$ .
An examination of the perturbation of spacetime-structure 
of the (relative) magnitude $10^{-24}\dots 10^{-26}$ as typical for 
the continous waves of pulsars must thus be done  
on scales which are large compared with the Planck length 
$L^*=1.616\cdot 10^{-35}\,$m, divided by this magnitude. This will be evidently 
not the case for an atom in the ground state or its first excited states. 

The following calculation examines the quantum 
mechanical system under consideration  
moving on a smooth spacetime-structure along geodesics during the induced         
transitions. A semi-classical calculation will 
be carried out  which makes use of a quantum system embedded in a 
classical space-time-manifold background. (For a complete description of the conditions used
see {\em Parker}
(\cite{parker}).)


In the first section 
the transition probabilities and absorption
cross sections for gravitational waves in quasi-hydrogenic systems are calculated. 
Emphasis will be laid on a {\em separation} of the dependence on fundamental 
constants and quantum numbers, because one is interested especially 
in the gain which might be obtained in higher (principal) quantum numbers.
In the  second section this gain  will be calculated explicitly with
an estimation of the upper limit for the relevant function of the quantum numbers.
The third section relates the attainable orders of magnitude for the transition 
probabilities with astrophysically realistic values of the wave amplitudes.
The validity of the implicit assumption `no magnetic field'  
in the derivation of the transition probabilities is discussed.
The fourth section is concerned with the 
fundamental constants which appear in the expressions derived 
in the second section. 
In the appendix 
  the `obligatory'  validity of Bohr's correspondence principle for the 
  case 
  of circular orbits is shown.
\section{Calculation of the absorption cross section for gravitational 
waves in 
quasi-hydrogenic systems}                                                       
In this section the transition between two fine structure levels with the respective 
quantum numbers\vspace{1.5mm}           
$$(n,l,j,m)                                                                    
\rightarrow (n',l',j',m')$$ 
in a quasihydrogenic system (an electron in the field of a point charge $Ze$) 
is considered. In the nonrelativistic  aproximation with  
respect to $e^-$-velocity ($Z\alpha \ll 1$) the interaction
operator  of the curvature of spacetime with the atom has the 
form (\cite{parker}):
\begin{equation}               
H_I=\frac12 m_ec^2R_{0l0m}x^lx^m\quad .
\end{equation}
This expression is quite as expected
a translation of the classical interaction energy expression for Newtonian systems
into quantum mechanics
via the rule of Jacobi. The coordinates $x^i$ are those of a Fermi-Normal-Coordinate
System centered at the core. The 
usual rigid core approximation is used in this expression. It can be shown however, that
the validity of all forthcoming relations holds                                
 if one lays the coordinate origin in the center of mass of the system
 and sets  $m_e$ equal the reduced mass $\mu=m_em_k/(m_e+m_k)$.

For the case
of a unpolarized wave one has then, following \cite{leen}:
\begin{equation}
H_I=-\sqrt{\frac{4\pi}{15}}m_e\omega 
\sqrt{4\pi S}      
\left( \frac{G}{c^3}\right) ^\frac{1}{2}r^2
i[Y_{2,2}(\vartheta,\varphi)-Y_{2,-2 
}(\vartheta                                                                     
,\varphi)]\cos( \omega t)\,\,\, ,\label{HI}  
\end{equation}   
with  $S$ being the energy flux density and  $\omega $ the frequency of the gravitational wave.
$Y_{l,m}$ are the spherical harmonics. 

The transition rate  will be now calculated
in first order Dirac perturbation theory.
The transition rate per unit time
between states with quantum numbers
$\gamma $,$\gamma $' ($\gamma =(nlj)$)  is obtained
by summing  the 
squared absolute matrix elements of the Fourier components of $H_I$
for negative and positive frequencies
over end states and then averaging over initial states.
Using the Wigner-Eckart theorem 
(\cite{biedenharn},\cite{sobelman}) and the usual eigenfunctions 
of the quasihydrogenic spherical-symmetric Coulomb-Problem
 the expression
\begin{eqnarray}                                                                
\lefteqn{\dot P^{(1)}_{\gamma \rightarrow \gamma '}                                      
=\frac{1}{2j+1} \sum_{m m'}\frac{2\pi }{\hbar ^2}\mid <                         
\gamma ' m'\mid H_I(\omega )\mid \gamma m>\mid ^2                                        
\delta (\omega -\omega _{\gamma \gamma '})} \nonumber \hspace{1.1em}\\ 
 & & =\frac{2\pi }{\hbar ^2}\frac{4\pi}{15}\frac{m_e^2G}{c^3}
 \omega^{2}4\pi S        
\nonumber\\
& &\qquad\times
\, \frac{1}{4}\frac{1}{2j+1}\sum_{m m'}                                                                  
\mid <n'j'l'm'\mid r^2(Y_{2,2}-Y_{2,-2})\mid njlm>\mid^2 \delta 
(\omega -       
\omega_{\gamma \gamma '})\nonumber\\                                            
 & & =\frac{4\pi ^2}{15}\frac{m_e^2G}{\hbar ^2c^3}\omega ^2S\left(                   
I_R^{n'l'nl}(2)\right) ^2\frac{1}{2j+1}\left|\left(\frac{1}{2}l'j'
\left\| C^{(2)}    \right\|                                                                      
\frac{1}{2}lj\right)\right|^2\delta (\omega -\omega _{\gamma \gamma '})
\,\,\, .\nonumber\\
\end{eqnarray}
results.
In this expression $S$ 
is the total intensity of the incoming gravitational radiation, 
independent of the polarization degree,
because it can easily be shown that the transition probability
is not altered when the radiation is polarized.
The radial integrals $I^{n'l'nl}_{R}(2)$
are defined
with the radial eigenfunctions $R_{nl}$
 of the Coulomb problem as
\[
I^{n'l'nl}_{R}(2)\equiv \int                                        
\limits_{0}^{\infty}dr\, r^4R_{n'l'}(r)R_{nl}(r)=<R_{n'l'}\mid r^2
\mid R_{nl}>     \quad .
\]
The so-called {\em reduced matrix elements} $(\frac12 l'j'\|C^{(2)}\|\frac12 lj)$, which are 
`reduced' from matrix elements  
of the operator                                                              
$C^l_m=\sqrt{4\pi/(5l+1)}\, Y_{lm}$, e.g. here
$C_m^2=\sqrt{{4\pi }/{5}}\, Y_{2m}$, appear in the following 
form
\[
C_{lj}^{l'j'}(2)\equiv
\frac{1}{2j+1}\left| \left(\frac{1}{2}l'j'
\left\| C^{(2)}   
\right\|\frac{1}{2}lj\right)\right|^2\,\, .
\]
The fundamental constants will in the following be absorbed into the constant
\[
A_Z\equiv \frac{4\pi ^2}{15}\frac{m_e^2G}{\hbar ^2c^3}a_0^4Z^{-4}=    
\frac{\pi}{15}(\lambda _e)^2\frac {G}{c^5}(\alpha Z)^{-4}\quad ,
\]
where $a_0={\hbar ^2}/{m_ee^2}$
is Bohr's radius, 
$\alpha ={e^2}/{\hbar c}$ the fine structure constant and
$\lambda _e={h}/{m_ec}$ the Compton-wavelength of the
electron.
With these definitions the transition probability takes the form
\begin{displaymath} 
\dot P^{(1)}_{\gamma \rightarrow \gamma '}=A_Z\omega         
^2S\, C_{lj}^{l'j'}(2) \left( \frac{I_{R}^{n'l'nl}}{(\frac{a_0}{Z})^2}
\right) ^2   
\delta (\omega - \omega _{\gamma \gamma '}) \,\,\, .
\end{displaymath}                    
This relation holds when the relevant states have
infinite lifetimes.
Realistically the transitions happen not only at one exact frequency 
(and then   
with a probability having no upper bound: {\it every} quantum  with 
$ \omega=  
\omega_{\gamma \gamma'} $ will be absorbed), instead the transitions 
will occur 
 within a total width                                                      
$\Gamma _{\gamma ,\gamma '}=\Gamma_ {\gamma }+\Gamma_                           
{\gamma '} $ ,
where $ \Gamma _{\gamma }=(\tau _{\gamma})^{-1}$ is the decay probability
and $\tau_{\gamma}$ the lifetime                                                       
of the excited state against electromagnetic decay (which is of course
by far the most important decay channel).
If one imprints a Lorentz profile on the transition probability
\begin{displaymath} 
b(\omega )d\omega =\frac{\Gamma /{2\pi }}{(\omega -    
\omega _{\gamma \gamma '})^2+\Gamma^2 /{4}}\quad ,
\end{displaymath}               
i.e. $ b(\omega )d\omega $ is the probability ($\int_{-\infty}^{\infty}        
 b(\omega )d\omega =1 $) for the absorption of a quantum with 
 frequency
$ \omega \dots \omega +d\omega $ making a transition 
$ \gamma \rightarrow
\gamma '$,                                                                       
one obtains for the transition probability per unit time and frequency        
interval:
\begin{equation}\frac{d\dot P^{(1)}_{\gamma \rightarrow \gamma '}}{d\omega}
=A_Z\omega_{\gamma \gamma '}^2S(\omega )C_{lj}^{l'j'}(2)\left(                 
\frac{I_{R}^{n'l'nl}}{(\frac{a_0}{Z})^2}\right) ^2\frac{{
\Gamma _{\gamma , \gamma '}}/{2\pi }}{(\omega -\omega _{\gamma \gamma '})^2+                       
{\Gamma^2_{\gamma ,  \gamma '}}/{4}}\,\,\, .\label{ddotP1}
\end{equation}                                                    
For the absorption cross section by influence of gravitational waves
one has
\begin{eqnarray}
\lefteqn{\sigma_{{\rm (abs)\gamma \rightarrow \gamma'}}(\omega)=\frac{\hbar \omega \, 
d\dot P^{(1)}_{\gamma   
\rightarrow \gamma '}(\omega)}{S(\omega)d\omega}}\nonumber\hspace{1.5em}\\
& & \cong A_Z
\omega_{\gamma \gamma '
}^3C_{lj}^{l'j'}(2)\left(                                                       
\frac{I_{R}^{n'l'nl}}{(\frac{a_0}{Z})^2}\right) ^2
\frac{{\Gamma _{\gamma , \gamma '}}/{2\pi }}{(\omega -\omega _{\gamma \gamma '})^2+                       
{\Gamma^2_{\gamma ,                                                        
\gamma '}}/{4}}\,\,\, ,
\end{eqnarray}                                                    
which is valid in the vicinity ($\mid \omega -                   
\omega_{\gamma \gamma '}\mid \ll \omega_{\gamma \gamma '}$) of the 
maximum. 

Assuming that the same relations hold for the relations between                      
spontaneous and induced emission- and                                            
absorption-probabilities for gravitational as for electromagnetic 
radiation, the Einstein relations  
will be valid. The probabilities $dw_{\hat k}/d\Omega$ 
per differential solid angle 
$d\Omega$ are related for unpolarized radiation  in the following way:
\begin{equation} g_{\gamma }\frac{dw^{{\rm abs}}_{\hat k}}{d\Omega}=
g_{\gamma '}      
\frac{dw^{{\rm ind}}_{\hat k}}{d\Omega}=g_{\gamma '}\frac{                            
dw^{{\rm sp}}_{\hat k}}{d\Omega}\frac{4\pi ^3c^2}{\hbar 
\omega _{\gamma \gamma '}^3}  
S_{\hat k}(\omega )\,\,\, . \end{equation}                                              
Here $S_{\hat k}(\omega )$ means the spectral intensity in direction $\hat k$ around 
$\Omega   
\dots \Omega+d\Omega$ and  $g_{\gamma}$ the degeneracy of the level with quantum 
numbers  $\gamma$.                                                                      

For a transition of width $\Gamma _{\gamma ,\gamma '}=\Gamma_ {\gamma }+
\Gamma_ 
{\gamma '} $ the spontaneous transition rate per unit time, frequency and 
solid 
angle interval is then                                        
\begin{displaymath}\frac{d^2w^{{\rm sp}}_{\hat k}}{d\Omega\, d\omega}                   
=\frac{g_{\gamma }}{g_{\gamma '}}\frac{\hbar 
\omega _{\gamma \gamma '}^3}       
{S_{\hat k}(\omega )4\pi ^3c^2}\frac{                                           
dw^{{\rm abs}}_{\hat k}}{d\Omega\, d\omega}\,\,\, .
\end{displaymath}                                                               
Using (\ref{ddotP1}) and integrating over 
all solid angles and frequencies (where
$dw^{abs}_{\hat k}/d\Omega\, d\omega = d\dot P^{(1)}_{\gamma \rightarrow 
\gamma'}/d\omega $
and $S_{\hat k}(\omega ) =S(\omega )$)
the total spontaneous transition rate 
for gravitational waves in   atomic transitions results:   
\begin{equation}\label{spont}\Gamma_{{\rm gr} ,\gamma '\rightarrow \gamma                            
}^{({\rm sp})}=\frac{1}{\tau_{{\rm gr} ,\gamma ' \rightarrow \gamma }^{({\rm sp})}}=
\frac{\hbar A_Z}
{\pi ^2c^2}\, \omega_{                                                             
\gamma \gamma '}^5\,\frac{g_{\gamma }}{g_{\gamma '}}\, C_{lj}^{l'j'}(2)
\left(
\frac{I_{R}^{n'l'nl}}{(\frac{a_0}{Z})^2}\right) ^2 \,\,\, .
\end{equation}                                                                  
We can define the quantity $\tau_{{\rm gr}, \gamma}^{({\rm sp})}$, which 
is the lifetime of  the
state $\gamma $ against spontaneous emission of gravitational waves:
\[
\tau_{{\rm gr}, \gamma}^{({\rm sp})}=\left[ \sum_{\gamma '} ( 
\tau_{{\rm gr} ,\gamma \rightarrow \gamma '}^{({\rm sp})})^{-1}\right] ^{-1} \]
(Where summation is over all 
 $\gamma '$ with $E_{\gamma}>E_{\gamma '}$.)

For the transition rate (\ref{ddotP1}) one may write (the superscript 
`(1)' in $d\dot P^{(1)}_{\gamma\rightarrow \gamma '}/d\omega$ 
is omitted from now on
and replaced by `(abs)'):
\begin{equation} \label{transratesimple}                 
\frac{d\dot P^{(abs)}_{\gamma \rightarrow \gamma '}}{d\omega}
=\frac{\pi ^2c^2}{\hbar }\,\omega_{                     
\gamma \gamma '}^{-3}\, \frac{g_{\gamma '}}{g_{\gamma }}\,
\Gamma_{{\rm gr} ,\gamma '\rightarrow \gamma }^{({\rm sp})}\, S(\omega )                                           
\frac{\Gamma _{\gamma ,\gamma '}/2\pi }     
{(\omega -\omega _{\gamma \gamma '})^2+
\Gamma^2_{\gamma , \gamma '}/4}\,\,\, .
\end{equation}      
The cross section can be expressed in the simple form
\begin{equation}          \label{sigmasimple}
\sigma_{({\rm abs}) \gamma \rightarrow \gamma '}(\omega )=2\pi c^2                    
\,\frac{g_{\gamma '}}{g_{\gamma                                                   
}}\,\omega _{\gamma \gamma '}^{-2}\,\eta_{\gamma ' \gamma }\frac{{ 
\Gamma ^2 _{\gamma ,                                                            
\gamma '}}/{4}}{(\omega -\omega _{\gamma \gamma '})^2+{
\Gamma^2_{\gamma ,   
\gamma '}}/{4}}\,\,\, .
\end{equation}                                                                  
With $\eta_{\gamma ' \gamma }$ the {\it branching ratio}              
defined as                                                                      
\begin{displaymath}                                                             
\eta_{\gamma ' \gamma }=\frac{\Gamma_{{\rm gr},\gamma ' \rightarrow \gamma }^{({\rm sp})}}   
{\Gamma _{\gamma ,\gamma '}}=\Gamma_{{\rm gr},\gamma '\rightarrow \gamma 
}^{({\rm sp})}\,\tau  
_{\gamma \gamma '}\,\,\, .                                                             
\end{displaymath}                                                               
$\eta_{\gamma '\gamma }$ represents the ratio of gravitational spontaneous 
emission 
probability in the transition $\gamma '\rightarrow \gamma                       
$ to                                                                            
electromagnetic (spontaneous) total emission probability                        
 (see definition
of $\Gamma _{\gamma ,\gamma '}=(\tau_{\gamma\gamma'})^{-1}$) for
the same transition. Noteworthy is the fact 
that the expression (\ref{sigmasimple}) above apart from geometrical        
factors is obtained for                                                         
{\em any} kind of detector (see \cite{weinberg} $ \S $10).   

Strictly spoken the expression above is the absorption cross section 
without consideration of the re-emission of gravitational waves, or in other words the 
total cross section:  $\sigma                                                       
_{{\rm abs}}=(1-\eta)\sigma_{{\rm tot}}$. But the scattering process of a gravitational wave 
at atomic systems is such an improbable one ($\sigma _{{\rm scatt}}=\eta                        
\sigma _{{\rm tot}} $ with $ \eta<10^{-50}$!), that it can be completely neglected. 

If one carries out the following {\it normalizations}\,\footnote{That 
they make sense,    
 will be shown in the next chapter: normalized quantities will depend 
 only on   
$\gamma ,\gamma '$.}  
\begin{eqnarray}
\omega _{\gamma \gamma '}&=&\left(\frac{2{\rm Ryd}}{\hbar}Z^2\right)
\tilde
\omega _{\gamma \gamma '}=\frac{m_ec^2(\alpha Z)^2}{\hbar}\tilde 
\omega _{\gamma
 \gamma '}=4.134 \cdot 10^{16}\mbox{sec}^{-1}Z^2\tilde \omega _{\gamma
 \gamma '}      \quad ,
\nonumber \\                                                                    
\tau_{\gamma \gamma '}&=&\left(\frac{{\rm Ryd}}{\hbar}\alpha ^3Z^4\right)^{-1}\tilde 
\tau_{\gamma
\gamma '}=\frac{2\hbar}{m_ec^2\alpha ^5Z^4}\tilde 
\tau_{\gamma \gamma '}
=1.245 \cdot 10^{-10}\mbox{sec}\, Z^{-4}\tilde \tau_{\gamma \gamma '} \quad ,
\nonumber \\         
I_{R}^{n'l'nl}(2)&=&\tilde I_{R}^{n'l'nl}(2)\left(\frac{a_{0}}{Z}\right)^2\,\,\, \quad ,
\end{eqnarray}
where normalized quantities have a tilde,
the branching ratio becomes
\begin{equation}                                                                
\eta_{\gamma ' \gamma }=\frac{8}{15}\frac{m_e^2G}{e^2}(\alpha Z)^2
\frac{g_{\gamma                                                                 
}}{g_{\gamma '}}\, C_{lj}^{l'j'}(2)\left( \tilde I_{R}^{n'l'nl}(2)
\right) ^2       
\tilde \omega _{                                                                
\gamma \gamma '}^5\tilde \tau_{\gamma                                           
\gamma '}\,\,\, .                                                                       
\end{equation}                                                                  
Here, in a quite natural way, the characteristic ratio                      
${m_e^2G}/{e^2}=2.4\cdot 10^{-43}$ appears, describing the 
relative         
magnitudes of electromagnetic and gravitational interaction (in the atom).                 

The maximum absorption cross section (\ref{sigmasimple}) becomes
\begin{equation}                                                                
\sigma_{({\rm abs})\gamma \rightarrow \gamma '}^{{\rm max}}(
\omega _{\gamma \gamma '})=     
\frac{16\pi}{15}\left( \frac{\hbar G}{c^3} \right) \alpha                       
^{-3}Z^{-2}\underbrace                                                          
{C_{lj}^{l'j'}(2)\left(                                                         
\tilde I_{R}^{n'l'nl}(2)\right) ^2 \tilde \omega _{\gamma \gamma '}^3
\tilde     
\tau_{\gamma \gamma '} }_{\equiv f(\gamma , \gamma ')}\,\,\, .
\end{equation}                                                                  
$({\hbar G}/{c^3})^{1/2}=L^*=1.616\cdot 10^{-35}\,$m  is the 
Planck length.                                                                 
Except the sets of quantum numbers $\gamma , \gamma '$ and the number $Z$   
the absorption cross section depends just on $L^*$ and the                     
fine structure constant                                                         
$\alpha $:
\begin{equation}\label{sigmaabs}                                                                
\sigma_{({\rm abs})\gamma \rightarrow \gamma '}^{{\rm max}}(
\omega _{\gamma \gamma '})=     
\frac{16\pi }{15}L^{*2}\alpha ^{-3}Z^{-2}f(\gamma ,\gamma ')\,\,\, .
\end{equation}                                                                  
Correspondingly  the expression (\ref{transratesimple})
for the transition rate is in 
resonance ($\omega =\omega _{\gamma \gamma '}$):                                        
\begin{equation}          \label{14}
\frac{d\dot P_{\gamma \rightarrow \gamma '}^{{\rm (abs)max}}}{d\omega}=
\frac{16\pi }{15}   
L^{*2}\alpha ^{-5}Z^{-4}\,\frac{S(\omega_{\gamma \gamma '})}{m_ec^2}  
\, \tilde  \omega_{\gamma                                                          
 \gamma '}^{-1}\, f(\gamma ,\gamma ')\,\,\, .                                         
\end{equation}                                                                  
For the case of nearly monochromatic waves, i.e. when the bandwidth of the 
radiation is much less than the width $\Gamma_{\gamma ,\gamma '}$                         
of the atomic transition, this expression transforms into
\begin{equation}              \label{15}
\dot P_{\gamma \rightarrow \gamma '}^{({\rm abs})max}=\frac{1}{30}\frac{c}{r_0}
\lbrack        
|A^{\times }|^2+|A^{+}|^2\rbrack\, \tilde  \omega_{\gamma                          
 \gamma '}f(\gamma ,\gamma ')\,\,\, ,
\end{equation}                                                                  
\begin{displaymath}                                                             
\quad \mbox{where}\qquad S=\int S(\omega )d\omega=\frac{c^3}{32\pi G}\,\omega_{\gamma \gamma '}^2
\lbrack   
|A^{\times }|^2+|A^{+}|^2 \rbrack  
\end{displaymath}                                                               
was used.
Here $r_0={e^2}/{m_ec^2}=2.817\cdot 10^{-15}\,$m is the  classical                  
electron radius and $A^{\times },A^{+}$ are the amplitudes of the two possible        
wave polarizations.                                                               
\section{Dependence of the absorption cross section 
and transition rates on the 
quantum numbers}  \label{depend}
For the function $f(\gamma , \gamma ')=f(nlj,n'l'j') $, defined as                              
\begin{displaymath}                                                             
f(\gamma , \gamma ')=C_{lj}^{l'j'}(2)\left(                                     
\tilde I_{R}^{n'l'nl}(2)\right) ^2 \tilde \omega _{\gamma \gamma '}^3
\tilde     
\tau_{\gamma \gamma '}   \quad ,
\end{displaymath}                                                               
an upper limit can be estimated for Rydberg states as follows:
\begin{itemize}
\item The normalized radial integrals $\tilde I_{R}^{n'l'nl}(2)$ can be 
estimated   
with                                                                            
Schwarz's inequality in the form                                                
\begin{displaymath}                                                             
\left(\int dr\, r^4R_{n'l'}(r)R_{nl}(r) \right) ^2\le \left( 
\int dr\, r^4R^{2}_{nl}  
\right)                                                                         
\left(\int dr\, r^4R_{n'l'}^{2}\right) =\bar{r^2}_{nl}\,\bar{r^2}_{n'l'}
\end{displaymath}                                                               
as                                                                              
\begin{displaymath}                                                             
\tilde I_{R}^{n'l'nl}(2)\equiv n^4i(n,l,n',l')\le n^4\left(                     
i(n',l',n',l')i(n,l,n,l)                                                        
\right)^{1/2}\equiv n^4i(\tilde n, \tilde l)\,\,\, .                                  
\end{displaymath}                                                               
$
i(n,l,n,l)=5/2+{1}/{2n^2}-{3l(l+1)}/{2n^2}
$ 
is easily 
calculated with  
 the well-known analytical result for the mean value  
 of $r^2$ in the state $(nl)$.   
\item For the normalized frequencies $\tilde \omega_{\gamma \gamma '}$ 
the familiar     
expression for rydberg states
 \[
 \tilde \omega_{\gamma \gamma '}\cong \Delta n/n^3\;, \; \mbox{where} \quad 
 \Delta n=n'-
n \qquad \Delta n\ll n\qquad \] 
is used.
\item The normalized total life time, calculated from 
$\tilde 
\tau_{\gamma , \gamma '}
=\left(\tilde 
\tau_{\gamma }{}^{-1}+\tilde \tau_{\gamma '}{}^{-1}\right) ^{-1}$, 
has,  following   
\cite{chang}, the approximate value (as an upper bound):
\begin{eqnarray*}
\lefteqn{\tilde \tau_{\gamma }=\tilde \tau_{nl}\le \frac34 
n^3(l+1/2)^2\quad ,}\hspace{-1.6em}\\ 
& & \tilde \tau_{\gamma \gamma '}=\tilde \tau_{nln'l'} 
\le \frac34 
n^3(l+1/2)^2       
\left( 1+\left(\frac{n}{n'}\right) ^3\left( \frac{l+1/2}{l'+1/2} 
\right)
^2\right)^{-1}\; .
\end{eqnarray*}
For Rydberg states  $(\Delta n\ll n)$ this simplifies to
\[
\tilde \tau_{\gamma \gamma '}=\tilde \tau_{nln'l'}
\le \frac34 
n^3(l+1/2)^2       
\left( 1+\left( \frac{l+1/2}{l'+1/2} \right)                                    
^2\right)^{-1}\; ,
\]
with a deviation from the real value of $\tilde \tau_{\gamma}$ 
which is,        
after \cite{chang}, less than 10 percent                                               
to lower values.
\item The quantities $ C_{lj}^{l'j'}(2)$, representing the `reduced 
matrix elements',
have, depending on the  transitions considered, 
nonvanishing values in the range $0\ldots 3/5$.
The exact values of $ C_{lj}^{l'j'}(2)$, paying attention to the 
selection rules
\[
\Delta j=0,\pm 1,\pm 2 
\quad ,\qquad \Delta l=0,\pm 2 \quad ,\qquad l+l'\geq 2    \quad ,
\]
are collected in table \ref{Ctab}. 
(Extracted from \cite{biedenharn}.)                
\end{itemize}
\renewcommand{\arraystretch}{1.2}
\begin{table}\caption{\label{Ctab}  ``Reduced matrix elements''}
\[\begin{array}{|r|c|c|}\hline                                                          
\multicolumn{3}{|c|}{\rule[-3mm]{0mm}{8mm} C_{lj}^{l'j'}(2)=}\\
\hline\Delta j= & j=l+1/2\quad j'=l'+1/2 &                                        
j=l+1/2 \quad j'=l'-1/2\\j'-j&\land\,\; j=l-1/2\quad j'=l'-1/2&\land \,\;
j=l-1/2
\quad j'=l'+1/2\\                                                              
\hline                                                                          
+2& \rule[-4mm]{0mm}{10mm}\frac{\textstyle 6(2j+3)(2j+5)}
{\textstyle 64(j+2)(j+1)}&0\\                                
\hline                                                                          
-2&\rule[-4mm]{0mm}{10mm}\frac{\textstyle 6(2j-1)(2j-3)}
{\textstyle 64j(j-1)}&0\\                                     
\hline                                                                          
+1&0&\rule[-4mm]{0mm}{10mm}\frac{\textstyle 6(2j+3)}
{\textstyle 32(j+2)(j+1)j}\\                                      
\hline                                                                          
-1&0&\rule[-4mm]{0mm}{10mm}\frac{\textstyle 6(2j-1)}
{\textstyle 32(j+1)j(j-1)}\\                                      
\hline                                                                          
0&\rule[-4mm]{0mm}{10mm}\frac{\textstyle (2j+3)(2j-1)}
{\textstyle 16(j+1)j}&0\\\hline                                       
\end{array}\]                                                                   
\end{table}

An upper bound for  $f$ is then:     
\begin{eqnarray}                                                                
\lefteqn{f(n,l,j,n',l',j')\le \frac{3}{4}C_{lj}^{l'j'}(2)\, i^2(\tilde n \tilde l)
\, (\Delta   
n)^3}\hspace{7em}\nonumber\\
& & \times\left(                                                                      
1+\left(\frac{n}{n'}\right) ^3\left( \frac{l+1/2}{l'+1/2} \right)^2
\right)^{-1} 
\left( n(l+1/2) \right)^2\,\, .                                                       
\end{eqnarray}                                                                  
This upper bound will reach the real value asymptotically for             
$n\rightarrow \infty $, $ l \simeq n-1$, i.e. for Rydberg atoms in 
nearly 
circular orbits, for the following reasons\,\footnote{`Fortunately' just 
these cases are the 
most interesting ones, because the most significant dependence on $l$             
(which leads to an increase in $l$!) is found                                    
in the quadratic increase of the lifetime with $l$.}:
\begin{enumerate}
\item The radial integrals for these `orbits' respectively for their wave          
functions with a small number of nodes $n_{r}=n-l-1$ are then nearly 
identical  
with the                                                                        
integrals for the mean value of $r^2$ $ \Longrightarrow i(\tilde n         
\tilde l)\cong                                                                  
1$.                                                                           
\item The upper bounds for the life times are then near the actual life 
times
(this holds for any $n$!).
\end{enumerate}
Therefore one has for $n\rightarrow \infty$ and $j\rightarrow \infty$
\begin{eqnarray}                                                                
\lefteqn{f(nlj,n'l'j')\cong \frac38 C_{lj}^{l'j'}(2)\, (\Delta n)^3\left( n(l+1/2) 
\right)^2}       
\nonumber\hspace{3.85em}\\ 
& &\cong 
\left\{\begin{array}{ll}
({\displaystyle 9}/{\displaystyle 64})\,
(\Delta n)^3\left( n(l+1/2) \right)^2 
&  \label{fkreis}              
\,\,\,(\Delta j=\pm 2)   \\
\,\,0 
\,\,\mbox{ (relative to } \Delta j=\pm 2) 
& \,\,\,(\Delta j=\pm 1)\\ 
({\displaystyle 3}/{\displaystyle 32})\,
(\Delta n)^3 \left( n(l+1/2) \right)^2 
& \,\,\,(\Delta j=0) \,\,\, . 
\end{array}\right.
\end{eqnarray}          
The maximum absorption cross section (\ref{sigmaabs})
may then be written for sufficiently high values of $l,l'$ in the asymptotical
form
\begin{equation} 
\sigma_{({\rm abs})\gamma \rightarrow \gamma '}^{{\rm max}}\cong                            
\frac{2\pi}{5}L^{*2}\alpha^{-                                                  
3}Z^{-2} \, C_{lj}^{l'j'}(2)\, (\Delta n)^3                                           
\left( n(l+1/2) \right)^2\,\, .                                                       
\end{equation}     
Similarly for transitions between nearly circular orbits the 
(approximative) identity 
\begin{equation}   
\frac{d\dot P_{\gamma \rightarrow \gamma '}^{{\rm (abs)max}}}{d\omega}\cong 
\frac{2\pi}{5}
L^{*2}\alpha^{-5}Z^{-4} \, C_{lj}^{l'j'}(2)\, \frac{S(\omega_{\gamma \gamma
'})}{m_ec^2}\, (\Delta n)^2 \left( n^{5/2}(l+1/2) \right)^2                        
\label{dwS}
\end{equation}
for the transition rate per unit time and frequency 
interval (\ref{14}) holds.                                                         
Finally, when the bandwidth of the incoming radiation is much smaller
than the total width of the transition regarded  (`monochromatic' radiation),
in the case of great $l,l'$, one obtains  
for the expression (\ref{15})
\begin{equation} \label{uebergangmonochrom}                                                            
\dot P_{\gamma \rightarrow \gamma '}^{{\rm (abs)max}}\cong \frac{1}{80}                    
\frac{c}{r_0}\lbrack                                                            
|A^{\times }|^2+|A^{+}|^2\rbrack \, C_{lj}^{l'j'}(2)\, (\Delta n)^4
\frac{(l+1/2)^2}{n}
\,\,\, .
\end{equation}                                                               
\section{Transition rates
in the field of astrophysical sources and the presence of magnetic fields}    \label{OoM}
A typical frequency of a monochromatic gravitational wave lies, 
in the case of 
the pulsar in the
Crab nebula, at $\omega=379.8\,$ Hz. (All values used in the following
were taken out of \cite{thorne} and publications cited therein.)
This corresponds to a transition 
with $\Delta n=1 $  of (approximately) $n$=47746 to                                            
$n'$=47747. (Here and in the rest of this section $Z=1$, where
an explicit value of $Z$ is required.) 
Clearly, one has to take care giving an        
integer $n$ (which is necessary for a transition to occur),
because the frequencies of the transitions in these regions 
of the principal quantum number are defined 
extremely sharp.    
The lifetime for $n=47746$ $l=47745$, for example (assuming that            
the calculations in \cite{chang} are still valid for such values of $n$), 
has the enormous 
value $7.35 \cdot 10^5\, $years! The condition of a monochromatic 
gravitational wave  would just {\em not} be fulfilled in the   
case of the Crab pulsar, because its          
lifetime is around 
$3\cdot 10^4$ years and the bandwidth of the emitted wave lies 
within a  corresponding region.
For exact calculations the fine structure must thus  be considered 
although it  vanishes with $n^{-1}$ relative to the Rydberg energies.                     
Typical values for $n$ may be set within a region $n=10^4\dots 10^5$                  
because it will be sufficient for an estimation neglecting for the 
moment that  the frequency will be exactly the right one.

One defines characteristic amplitudes $|A|$ 
by the relation
\[
|A|=\left(\frac{16\pi G}{c^3}\frac{S}{\omega        
^2}\right) ^{1/2}\quad , 
\]
where $|A|=|A^+|=|A^{\times}|$ and  $S$ is the total wave flux.
For continous waves  emitted by pulsars the predicted values typically
are $ |A|=10^{-25}\dots 10^{-26}$ (Crab: $|A|=10^{-24}\ldots 10^{-27}$).  
According to the relation (\ref{uebergangmonochrom}) from the 
preceding section  one has as a maximal transition probability
for transitions with                                             
$\Delta n=1 \, ,\, \Delta j=2 $ between nearly circular orbits 
($C_{lj}^{l'j'}(2)\simeq 3/8$) of
\begin{equation}\label{monochromesimple}
\dot P^{\rm (abs)max}_{n\rightarrow n+1}\cong 10^{21} \mbox{sec}^{-1}|A|^2n
\qquad (l,l'\simeq n)\,\, \, ,
\end{equation}
where the idealizations described above are assumed.
For $n=10^5$, $|A|=10^{-25}$  
the transition probability has the value
$\dot P\cong 10^{-24}\mbox{sec}^{-1}$. Under very optimistic assumptions 
with $n=10^4$, $|A|=10^{-20}$ one would get $\dot P\cong 10^{-15}\mbox{sec}^{-1}$.

Concerning a possible detection of the waves one has to consider 
first the enormous difficulties which have to be solved
trying to hold just one of these Rydberg      
atoms with $n\cong 5\cdot 10^4$ stable (they have the size of a soccer ball!),
then has to estimate the number of atoms with the principal quantum numbers 
required to reach a detection, and at last
one may not  forget the 
extremely restrictive conditions for the monochromatic character of the radiation.
(For the latter example above the number of atoms required for three
transition events per year would be $10^8$.)

Even if a solution of these difficulties
would seem to be not totally out of reach, there is at least
one serious obstacle which makes a practial realization especially on Earth
extremely difficult.
The bandwidth of the gravitational wave detector for the Crab Pulsar is
$\Delta f /f \approx (60.45 {\rm Hz}\times 3.67\cdot 10^5{\rm years})^{-1}
=1.4\cdot 10^{-15}$.
The Doppler shift caused by the Earth's rotation is $\Delta f/f
\approx 10^{-6}$ with a diurnal variation. The
detector will consequently be within the bandwidth of the radiation 
only a fraction of about $10^{-9}$ per day and for just about 0.1\,msec. 
On the other hand, a gravitational wave detector in {\em space} would have 
to maintain a relative line of sight velocity variation 
to the Crab pulsar smaller than $4.2\cdot 10^{-7}$m\,sec$^{-1}$. \\

We turn now to the question of
the validity of the calculated transition probabilities between
adjacent principal quantum numbers in the presence of magnetic fields.
To speak of a transition $n\rightarrow n+1$ makes sense as long as 
$n$ makes sense, i.e. as long as $n$ is a `good' quantum number.
In the presence of a magnetic field (constant and in $z$-direction), 
$n$ ceases to be a good
quantum number whenever adjacent energy levels with constant magnetic
quantum number $m$ but different principal quantum number begin 
to mix. 
A good estimation for an upper bound of the magnetic field in the
limit $n\rightarrow \infty$  
can be obtained when one sets the semiclassical diamagnetic energy  
for circular orbits $E_{\rm dia}=(1/8)r_0 a_0^2\!<\!x^2+y^2\!>\simeq
(1/8)r_0 a_0^2 \!<\!r^2\!>\simeq (1/8)\, r_0 a_0^2 Z^{-2}B^2 n^4$ 
($n\rightarrow\infty,\; l=n-1\approx n$) equal the Bohr energy 
$\hbar \omega_n =2\,{\rm Ryd}Z^2/n^3$. (The paramagnetic 
part gives the same contribution for every $n$ if  $m$ is constant.)  
This leads to $B\simeq\sqrt8 Z^2 a_0^{-1}(2\,{\rm Ryd}/r_0)^{1/2}\, n^{-3.5}=
6.6\cdot 10^9 Z^2\, n^{-3.5}\,{\rm Gauss}$.

The conditions set by this upper bound of $B$ for $n$ to remain a 
good quantum number are extremely stringent  
for the high principal quantum numbers discussed. For instance, 
setting $n=1000$, $B$ has already to be smaller
than 0.2\,Gauss, which is in the order of the Earth's magnetic field.
For $n=47746$, the (approximate) quantum number which is required for the frequency  
of the Crab pulsar, the value of $B$ is $2.8\cdot 10^{-7}$\,Gauss.
(One has to be reminded that the order of magnitude of the 
cosmic background field is $B\approx 10^{-6}$\,Gauss!)

This sheds light on the highly idealized character of the assumption 
`no magnetic field' for any calculation dealing with Rydberg atoms
which have $n\gg 1000$ and the extreme shielding requirements that 
one would have to meet keeping them `free' of magnetic fields.

\section{Fundamental constants}             
The dependences on fundamental  
constants, which were obtained in the calculations by `normalizing'
the relevant quantities, find their 
natural explanation in the assumption of the validity  of the equivalence
principle, which was made throughout the whole calculation. 
Correspondingly a simple geometric explanation of the mathematical expressions 
at least in a semiclassical limit may be used. 

In equation (\ref{sigmaabs}) for the maximum absorption  
cross section                                         
\begin{displaymath}                                                             
\sigma_{({\rm abs})\gamma \rightarrow \gamma '}^{{\rm max}}(\omega _{\gamma 
\gamma '})=     
\frac{16\pi }{15}L^{*2}\alpha ^{-3}Z^{-2}f(\gamma ,\gamma ')\,\,\, 
\end{displaymath}                                                               
one recognizes that the  absorption cross section is independent of a specific particle 
property as the    
mass-charge ratio (or the mass alone), in complete contrast to the 
analogous case of the absorption
of electromagnetic waves                                                                    
(in the fine structure constant $\alpha$, although, the elementary charge $e$  
is contained). For any comparable electrically bound system (with a 
stable ground 
state into which excited states decay electromagnetically) one 
 obtains         
the same cross section. In the electromagnetic theory, $G$ corresponds to 
$e^2/{m^2}$:                                     
\begin{displaymath}                                                             
L^{*2}=\frac{\hbar G}{c^3} \Longleftrightarrow \left( \frac{e^2}
{m_{e}c^2}      
\right) ^2                                                                      
\frac{\hbar c}{e^2}=r_0^2/\alpha=\frac{3}{8\pi}
\sigma_{\mbox{\tiny Thomson}}/\alpha\,\,\, .
\end{displaymath}                                                               
Now the classical particle radius of the classical electromagnetic theory at least in 
principle may be varied freely, depending on the particle which one takes 
`at hand'.
The {\it effective                                             
gravitoelectric (classical) particle radius} 
$L^{*}(\times \sqrt{ \alpha})=\left( Ge^2/c^4 \right)^{1/2}$
(if one uses the analogy above to the electromagnetic case)
is, in contrast,  fixed\,\footnote{
For considerations in a similar context see also Smolin \cite{smolin}.}.
In the theory of the interaction of 
Gravitation and         
electrically bound (atomic) systems this particle radius 
is an intrinsic property of the 
particle and   
independent of the mass of this particle. It is given by the nature 
constants $G, e$ and $c$. The `particle radius' defined in the way above must
not be confused  
with the particle radius defined analogous to the electromagnetic theory via 
the defining relation 
{\em potential self energy equals rest energy}, which leads to
$r_{0\,gr}=m_eG/c^2$.     
This radius depends on the mass and is a radius which is defined 
solely in Gravitation theory.

Another quite suggestive quantity can be formed with the ratio of the 
absorption
cross section for  gravitational waves
to the `real' cross section $\bar \sigma=4\pi    
<r^2>_{nl}$ in the state $(nl)$. One obtains with                               
$<r^2>_{nl}=1/2\left[ 5n^2+1-3l(l+1)\right] n^2a_{0}^{2}/Z^2$ 
and equation (\ref{sigmaabs}):                                                                 
\begin{displaymath}                                                             
\frac{\sigma_{{\rm (abs)}\gamma\rightarrow\gamma'}^{{\rm max}}}{\bar \sigma}=\frac{8}{15}
\frac{G}{\left(e/m_e                                                                   
\right)^2}\frac{f(\gamma , \gamma ')}{\left[ 5n^2+1-3l(l+1)\right] n^2}\quad .
\end{displaymath}                                                               
The limit $n\rightarrow \infty$ of this expression 
for transitions with $\Delta j=\pm 2$ 
between nearly circular orbits is\,\footnote{  
See first line of (\ref{fkreis}): $\Delta j=\pm  2\, ,
\quad l\cong n-1\, , \quad \Delta n =1\quad\rightarrow \quad 
f(\gamma \gamma ') =(9/{64})\, n^4$.}                                                           
\begin{displaymath}                                                             
\frac{\sigma_{{\rm (abs)}}^{{\rm max}}}{\bar \sigma}\cong \frac{3}{40} 
\frac{G}{\left(e/m_e         
\right)^2}\quad .                                        
\end{displaymath}                                                               
The ratio of maximal absorption cross section to `real' cross section 
of the highly excited atom has an approximate magnitude which  
has a quite simple explanation. 
Its value is proportional to the characteristic ratio of
gravitational to electromagnetic interaction (in the atom) and is 
a classical quantity (no $\hbar$!).
This result is not astonishing, because, especially on a graphic 
ground, the ratio for the electromagnetic case will be quite clearly in order of
magnitude 1 and is reduced for the interaction process
considered here by the
factor   ${m_e^2G}/{e^2}$: The gravitational wave `sees' just (an approximate)              
 part ${m_e^2G}/{e^2}$ of the                                               
atom.

The independence of  the absorption cross section on the mass  can be 
understood in   
terms of the equivalence principle. The absorption cross section is a 
result of the   
particle's `orbit' and the `shape' of this `orbit' under the influence 
of the   
gravitational wave (`orbit' and `shape' in a classical sense), 
and this orbit is, following the equivalence principle, independent of the particle's mass, 
so far a purely 
gravitational origin of the                                                                  
scattering mechanism is concerned. In so far an experimental 
verification     
would provide a test for the (atomic?) scales where the equivalence 
principle
 remains valid.
 
\section{Conclusion} 
In view of the resulting  magnitudes calculated in section \ref{OoM}
it can be stated that even the 
possibility of a detection of gravitational waves by resonant interaction
with highly excited atoms can be excluded 
under physical and practical   
realistic circumstances unless a source of monochromatical waves  with              
an unexpected and extraordinary combination of high frequency and 
intensity exists, which seems not very likely.

Besides severe practical difficulties the crux is obviously the slow increase 
of the transition rates with 
principal quantum number only proportional to $n$ (equation 
(\ref{monochromesimple})), when one is assuming a constant amplitude of the 
periodic spacetime-perturbation $|A|$ with $n$ and then also $\omega (n)$. For 
the (unrealistic) assumption of constant energy flux $S$ 
($S\sim \omega^2 \sim 1/n^6 $!) one would  have a very fast  
increase with $n^7$ (equation (\ref{dwS})). 

The rapidly growing effect of interaction strength and sharpness 
of the energy levels with rising $n$ gives an enhancement  of the 
transition  probabilities which is not sufficient.
The reason is that the energy of the relevant quanta to be absorbed 
is decreasing too fast with principal quantum number ($\propto n^{-3}$)
and that the radiation is of quadrupole nature, which strengthens 
the influence of the $n$-dependence of this energy.

\bigskip

\noindent{\bf Appendix. Classical limit}
\bigskip

 
 The power emitted in form of gravitational radiation by an electron 
 circulating 
 on an radius $r$ with angular frequency $\Omega$ is 
 (see e.g. \cite{weinberg}):
\begin{displaymath}
P(2\Omega)=\frac{32}{5}\frac{Gm_e^2}{c^5}\,\Omega^6\, r^4\,\,\, .
\end{displaymath}
This emission is at twice the frequency $\Omega$. Expressed in the 
emitted 
frequency $\omega=2\Omega$ one has for $P$ 
\begin{displaymath}
P(\omega)=\frac{1}{10}\frac{Gm_e^2}{c^5}\,\omega^6\, r^4\,\,\, .
\end{displaymath}
The `quantum mechanical' transition rate calculated from this 
classical
 result becomes
\begin{displaymath}
\Gamma=P/\hbar\omega=\frac{1}{10}\frac{Gm_e^2}{\hbar c^5}
\, \omega^5\, r^4\,\,\, .
\end{displaymath}
The quantum mechanical total spontaneous transition rate is, 
from equation (\ref{spont}):
\begin{displaymath}\Gamma_{{\rm gr}\gamma '\rightarrow \gamma                            
}^{({\rm sp})}= \frac{4}{15}\frac{Gm_e^2}{\hbar c^5} 
\omega_{                                                             
\, \gamma \gamma '}^5\, \frac{g_{\gamma }}{g_{\gamma '}}
\, C_{lj}^{l'j'}(2)\left(I_{R}^{n'l'nl}\right)^2\,\, .                             
\end{displaymath}                                                                  
The classical limit consists in letting $n\rightarrow \infty$ in
$I_{R}^{n'l'nl}$ and 
$j\rightarrow \infty$ (circular orbits!) in 
$ (g_{\gamma }/g_{\gamma '})C_{lj}^{l'j'}(2) $. 
The result for this limit is obtained from section 2   as
 \begin{displaymath}    
 \frac{g_{\gamma }}{g_{\gamma '}}C_{lj}^{l'j'}(2)
 \stackrel{j\rightarrow \infty}{\longrightarrow } \frac{3}{8}
 \quad (\Delta j=\pm 2)\qquad 
 \left( I_{R}^{n'l'nl}\right) ^2
 \stackrel{n\rightarrow \infty}{\longrightarrow} 
 \left(\frac{a_0}{Z}\right)^4n^8\,\,\, .
 \end{displaymath}    
 With $n^2a_0/Z\equiv r$, reflecting the fact that the width of 
 the probability 
 density distribution vanishes in the limit $n\rightarrow \infty$, the 
 result for the `classical' transition probability
 above is reproduced, showing clearly the validity of the correspondence 
 principle. 


\section*{References} 
\begin{thebibliography}{99}
\bibitem{leen} Leen T. K., Parker L. and Pimentel L. O. 1983 
{\em Remote Quantum Mechanical Detection of Gravitational Radiation} 
Gen. Rel. Grav. {\bf 15} 761.
\bibitem{gill} Gill E., Wunner G., Soffel M. and Ruder H. 1987 
{\em On hydrogen-like atoms in strong gravitational fields} 
Class. Quantum Grav. {\bf 4} 1031.
\bibitem{pinto} Pinto F. 1993 
{\em Rydberg Atoms in Curved Space-Time} 
Phys. Rev. Lett. {\bf 70} 3839.
\bibitem{parker} Parker L. 1980   {\em One-electron atom as a probe 
of spacetime curvature} 
Phys. Rev. D {\bf 22} 1922.  
\bibitem{biedenharn} Biedenharn L. C., Louck J. D.  1981 
Angular Momentum in Quantum Physics 
{\em Encylopedia of Mathematics and its 
Applications} {\bf 8}  (Addison Wesley).                                                     
\bibitem{sobelman} Sobelman I. I. 1979 {\em Atomic Spectra and 
Radiative Transitions} 
Springer Series in  Chemical Physics {\bf 1}) (Springer).                                                   
\bibitem{weinberg} Weinberg S.  1972 {\em Gravitation and Cosmology} 
(Wiley and Sons).
\bibitem{chang} Chang E. S. 1985 
{\em Radiative lifetime of hydrogenic and quasihydrogenic atoms} 
Phys. Rev. A {\bf 31} 495.               
\bibitem{thorne} Thorne K. S. 1987 {\em Three hundred years of Gravitation} 
edited by S. W. Hawking, W. Israel (Cambridge University Press).
\bibitem{smolin} Smolin L. 1985 
{\em On the Intrinsic Entropy of the Gravitational Field} 
Gen. Rel. Grav. {\bf 17} 417.
\end{thebibliography}
\end{document}